\documentstyle[preprint,epsfig,aps]{revtex}

\begin{document}
\hfill IFT-P.005/98

\hfill IFUSP/p-1293
\vskip 0.6cm

\centerline{\bf Gamma Ray Bursts as seen by a Giant Air Shower Array}

\centerline{C.O. Escobar$^{1,2}$, P.L. Da Silva$^2$ and R.A. V\'azquez$^{2,3}$}

\centerline{\it $^1$ Instituto de F\'{\i}sica ''Gleb Wataghim''}
\centerline{\it Universidade Estadual de Campinas, Unicamp}
\centerline{\it 13083-970 -- Campinas, S. P. }
\centerline{\it Brazil}

\centerline{\it $^2$ Instituto de F\'{\i}sica, Universidade de S\~ao Paulo,}
\centerline{\it Caixa Postal 66318 }
\centerline{\it 05315-970 -- S\~ao Paulo, S. P. }
\centerline{\it Brazil}

\centerline{\it $^3$ Instituto de F\'{\i}sica Te\'orica,}
\centerline{\it Universidade Estadual Paulista}
\centerline{\it Rua Pamplona, 145}
\centerline{01405-900, S\~ao Paulo, S.P.}
\centerline{\it Brazil}

\begin{abstract}
The potentiality of a Giant Shower Array to low energy gamma rays from gamma
ray bursts is discussed. Effective areas are calculated for different scenarios
and the results are encouraging. If gamma ray bursts have a spectrum which
continues in the high energy gamma ray region, the Pierre Auger Observatory
will be  able to detect it.
\end{abstract}

\section{Introduction}

Gamma Ray Bursts (GRB) are one of the most intriguing mysteries in the Universe
and models abound trying to explain them. A recent review of models is given in
ref. \cite{Rees_1}. In ref. \cite{Nemiroff_1} a compilation of more than 100
models is given. A complete phenomenological review with many references is
given in ref \cite{Fishman_1}.

Recently, thanks to the accurate measurement of the position of a GRB by the
BeppoSAX satellite \cite{Costa} it was possible to measure a source 
counterpart in other frequencies (X rays, optical and radio bands) and a 
measurement of  the redshift \cite{Metzger} was given for the GRB970228 (0.86
$\le z \le $ 2.3) implying a truly cosmological distance. 

In this respect it will be most interesting to measure the flux of high energy
photons from gamma rays burst. We should remember that one does not expect
high energy photons if GRB are cosmological. Even a 100 GeV photon flux would
suffer a considerable attenuation for cosmological distances ($\sim $ Gpc).
Therefore, looking for high energy photons would be an invaluable tool in
discriminating cosmological scenarios from extended halo or mixed models. 

In this letter we discuss the potential of a giant surface array, specifically
the Pierre Auger Project \cite{Auger} to measure bursts fluxes of low energy (
GeV -- TeV) photons. Studies of this type have been done for a variety of
detectors and limits have been put by different collaborations, see for
instance \cite{EASTOP,Cygnus}. All present results are negative and only upper
limits could be put on fluxes on high energy. An interesting result is,
however, the 10 $\sigma$ candidate event detected by EAS-TOP \cite{EASTOP}.
Although the event could not be associated to any GRB's this could be an
indication of the ''delayed phenomena'' \cite{Waxman_1}. In this work, the
acceleration of the  highest energy cosmic rays is related to GRB. Delayed high
energy photons appear naturally as a consequence of the propagation and
cascading of the cosmic rays through the photon (CMBR and IR) fog. The delay
time depends on the intergalactic magnetic structure and could be of the order
of hours to days (or even years).

Recently, a work similar to ours was written by DuVernois and Beatty
\cite{DuVernois_1} were calculation of the effective area for the Auger
detector was done. However, our results indicate that their effective area is
overestimated by a factor of 15 with respect to us. We have been unable to
trace back the cause of discrepancy.

\section{Simulation}
\label{simulation}

We have run several thousands photon initiated showers for different energies
using the program Aires \cite{Aires}. Threshold energies for both muons and
e.m. particles have been set above the respective threshold for Cerenkov light
production in water. The ground depth was set to 850 gr/cm$^2$ \cite{Auger}. In
table \ref{table1} we can see some parameters of the run. The thinning level is
always chosen so as to guarantee that all particles in the shower are kept
until they reach the threshold value. 

\begin{figure}
\epsfxsize=10cm
\begin{center}
\mbox{\epsfig{file=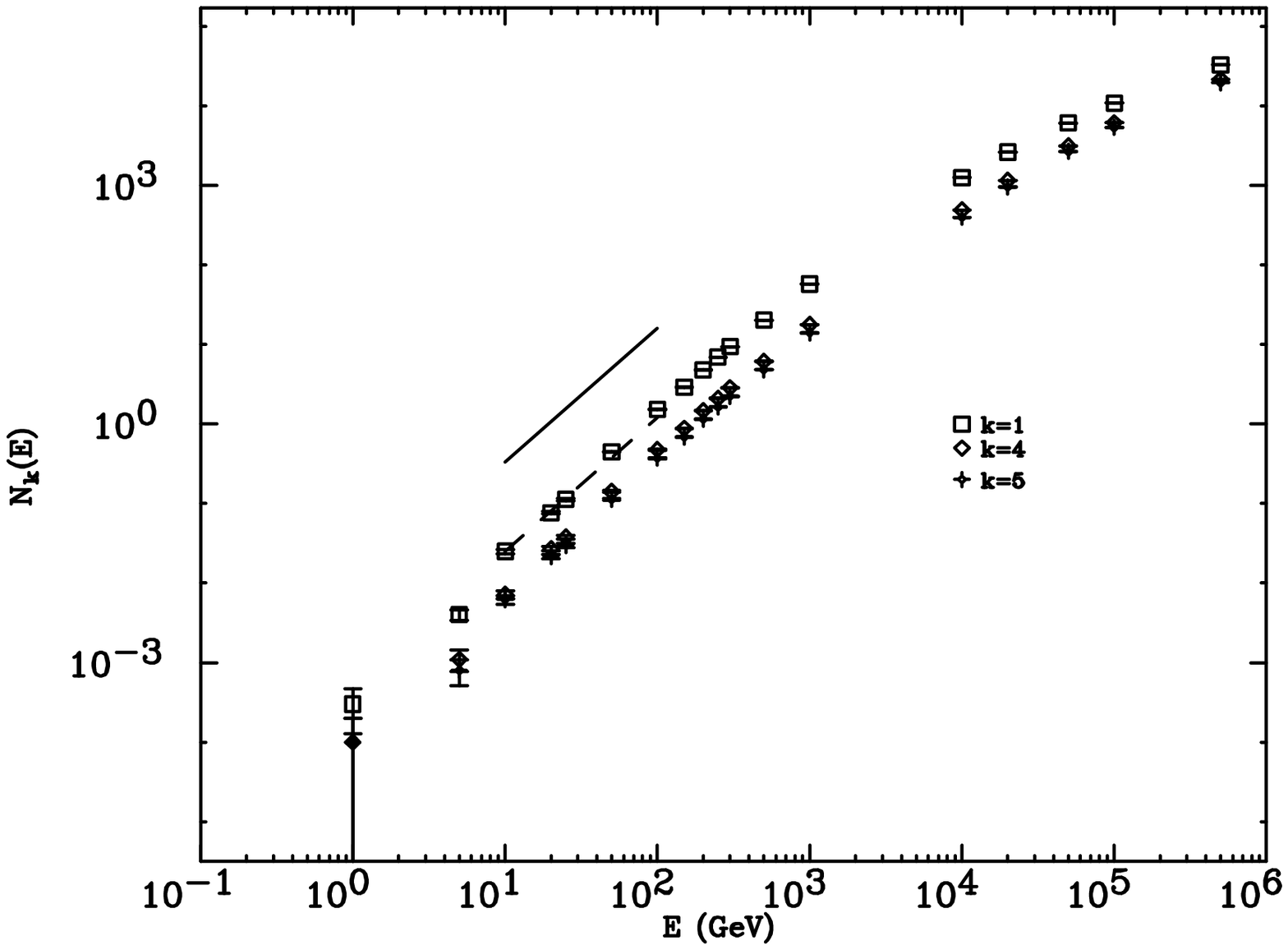}}
\end{center}
\caption{Number of triggers as a function of photon energy and $k$ (see text)
for photons arriving vertically. Also shown is the parametrization of DuVernois
and Beatty (continuous line) and the parametrization of EASTOP (dashed line).}
\label{fig1}
\end{figure}

For every shower we have done the following simulation. At ground level an
infinite grid is built. Each cell in the grid has an area of $a= 10 $ m$^2$
equal to the area of the detectors in the Auger Project. For each shower we
compute the number of particles that reach each cell at the ground. If the
number of particles is above a given number, $k$, we will count it as a
trigger. Results of the simulation are given in the table \ref{table1}. Let
$N_k(E)$ be the number of triggers for showers of energy $E$. Then a convenient
parametrization is given by:
\begin{equation}
N_k(E)= N_0(k) \; E^{1.68},
\end{equation}
where $N_0(k)$ depends on $k$ but it is independent of the energy. In
fig.\ref{fig1} we can see the result of our calculation of $N_k$ for different
values of $k$ as a function of energy. Also plot is the above parametrization. 
We can see that for medium energies ( 10 $< E < $ 100 GeV) the parametrization
is good. At high energies it overestimates the values of $N_k$.

Given the value of $N_k$ we can calculate the effective area for detecting low
energy showers as follows. A shower of some given energy will have an
''effective area'' given by:
\begin{equation}
A_{S}(E)= a \; N_k(E),
\end{equation}
where $a$ is the area of each detector.
This expression for the effective area reflects the low energy character of the
photon initiated shower which has shower maximum position much higher than the
ground level, giving a surface distribution of particles bearing no relation to
the initial shower axis direction. 

For an array of detectors, we have that the probability of a shower to trigger
a detector is given by the ratio of the  shower area by the inter detector area
{\it i.e.}:
\begin{equation}
P(E) = \frac{A_{S}(E)}{l^2},
\end{equation}
where $l$ is the separation distance between detectors. So the total effective
area will be:
\begin{equation}
A_{\mbox{eff}}(E) = A_T \; P(E),
\end{equation}
where $A_T \sim N_D l^2 \sim 3000 $ km$^2$ is the total area covered by the
detector and $N_D$ is the number of detectors. Finally we get:
\begin{equation}
A_{\mbox{eff}}(E) = a \; N_D \; N_k(E).
\end{equation}
From fig. \ref{fig1} we can estimate the effective area for the Auger Project.
At $E \sim$ 100 GeV we have $A_{\mbox{eff}} \sim a \; N_D \sim 1.6 \; 10^4$
m$^2$. Even at so low energies as $E \sim $ 10 GeV we have an effective area of
$A_{\mbox{eff}} \sim 16 $ m$^2$, {\it i.e.} bigger than the effective area of
EGRET \cite{EGRET}. This result is in contradiction with the  result in
ref.\cite{DuVernois_1}. In fig. \ref{fig1} we show their result as a continuous
line. We can see that their result is $\sim$ 15 times higher than ours even in
the most optimistic scenario of $k=1$. In comparison we also show the EAS-TOP
result \cite{EASTOP} which agrees with ours. This should be expected since the
altitude for EAS-TOP is similar to the projected Pierre Auger (850~gr/cm$^2$)
altitude and the area of each detector is equal (10 m$^2$). We conclude that
our results are correct.

Due to the large distance between detectors for the Auger Project ($l \sim 1.5
$ km) the array always have to operate in single counting mode. Even at the
highest energies ($\sim$ 100 TeV) the shower will certainly trigger on at most
one detector and no correlation on neighbour detectors should be expected. This
is in contrast to smaller arrays where the small distance between detector
allows to measure correlation between neighbour detectors see for instance
\cite{EASTOP}.

For the angular dependence we have simulated 1000 showers at fixed energies 
and at different angles. In table \ref{table2} we show the result of the
simulation. We can parametrize the results in the form:
\begin{equation}
N_k(\theta)= N_0 \; \cos(\theta)^\alpha,
\end{equation}
with $\alpha \sim 9$ which agrees with previous results
\cite{DuVernois_1,EASTOP}. We assume that this dependence is valid for other
energies. With the result of our calculation we are able to calculate the
effective area for GRB with arbitrary spectrum and for arbitrary arrival
direction. 

\begin{table}
\begin{displaymath}
\begin{array}{||c|c|c|c||}
\hline
\hline
\mbox{Energy (GeV)} & N_{sh} & N_{k=1} & N_{k=5} \\
\hline
1 & 10^4 & 3 & 1 \\
5 & 10^4 & 40 & 8 \\
10 & 10^4 & 250 & 62 \\
20 & 10^4 & 767 & 216 \\
25 & 10^4 & 1.12 \; 10^3 & 297 \\
50 & 10^4 & 4.45 \; 10^3  & 1.13 \; 10^3 \\
100 & 10^4 & 1.52 \; 10^4  & 3.72 \; 10^3 \\
150 & 10^4 & 2.89 \; 10^4  & 6.88 \; 10^3 \\
200 & 10^4 & 4.79 \; 10^4 & 1.16 \; 10^4 \\
250 & 10^4 & 6.91 \; 10^4 & 1.64 \; 10^4 \\
300 & 10^4 & 9.36 \; 10^4  & 2.23 \; 10^4 \\
500 & 10^4 & 2.02 \; 10^5 & 4.80 \; 10^4 \\
\hline
\mbox{Energy (TeV)} & N_{sh} & N_{k=1} & N_{k=5} \\
\hline
1 & 10^4 & 5.75 \; 10^5 & 1.39 \; 10^5 \\
10 & 5 \; 10^3 & 6.27 \; 10^6 & 1.98 \; 10^6 \\
20 & 10^3 & 2.61 \; 10^6  & 9.53 \; 10^5 \\
50 & 10^2 & 6.09 \; 10^5 & 2.67 \; 10^5 \\
100 & 10^2 & 1.09 \; 10^6 & 5.35\; 10^5 \\
500 & 70 & 2.29 \; 10^6 & 1.37 \; 10^6 \\
\hline
\hline
\end{array}
\end{displaymath}
\caption{Shower simulation parameters.}
\label{table1}
\end{table}
\begin{table}
\begin{displaymath}
\begin{array}{||c|c|c|c|c||}
\hline
\hline
\mbox{Energy = 100 GeV} & & & & \\
\hline
\theta & \cos(\theta) & N_{sh} & N_{k=1} & N_{k=5} \\
\hline
0 & 1 & 10^4 & 15234 & 3718 \\
10 & 0.985 & 10^3 & 1295 & 362 \\
20 & 0.949 & 10^3 & 786 & 196 \\
30 & 0.866 & 10^3 & 334 & 90 \\
40 & 0.766 & 10^3 & 75 & 19 \\
\hline
\hline
\end{array}
\end{displaymath}
\caption{Angular dependence on trigger number.}
\label{table2}
\end{table}

Let's assume a burst with a spectrum $dN/dE= \Phi_0 (E/E_0)^{-\gamma} $ 
between  $E_{\mbox{max}}$ and $E_{\mbox{min}}$ occurs during a time $T$. Then
the number of triggers in excess observed will be
\begin{equation}
S = T \; \int_{E_{\mbox{min}}}^{E_{\mbox{max}}} dE \; \Phi_0
\frac{E}{E_0}^{-\gamma} \; A_{\mbox{eff}}(E,\theta).
\end{equation}
In the same time the number of background triggers will be $N = \nu T$,
where $\nu$ is the background trigger rate. And the statistical significance is
given by:
\begin{equation}
n_\sigma = \frac{S}{\sqrt N} = \frac{a \; \Phi_0 \; T \; N_D \; 
\zeta(\gamma,\theta)}{\sqrt{T \; \nu \; N_D}}
\end{equation}
where $\zeta(\gamma,\theta)$ is given by
\begin{equation}
\zeta(\gamma,\theta)= \int_{E_{\mbox{min}}}^{E_{\mbox{max}}} dE \; E^{-\gamma}
N_k(E,\theta).
\end{equation}
Therefore a limit with a $n_\sigma$ confidence level can be obtained from no
observation for fluxes bigger than
\begin{equation}
\Phi_0 = \frac{n_\sigma}{a} \sqrt \frac{\nu}{T N_D}
\; \frac{1}{\zeta(\gamma,\theta)}.
\end{equation}

For the Auger detector the single counting ratio will be about 2.5 kHz and the
integration time can be of order 1 s, although given the uncertainty in the
time profile for high energy photons, specially if the delayed phenomena is
general, this time should be refined. Thus we get,  assuming a spectrum index
of 1.5 in the region of 1 GeV to 1 TeV:
\begin{equation}
\Phi_0 = 5.1 \times 10^{-6} n_\sigma \; \frac{\mbox{ph.}}{\mbox{cm}^2 \mbox{s
GeV}}. 
\end{equation}
We should use $n_\sigma \ge 10$ in order to avoid accidental triggers. With
this values we would get upper limit fluxes which are competitive with actual
Cerenkov detectors, see for instance ref. \cite{Baring_1}.

The present technique will not be able to give any indication of the arrival
direction of these photons, contrary to Cerenkov detectors, therefore the
actual detection will be possible only on the basis of timing considerations
and correlations with other experiments. An advantage, however, is that it is
sensitive to a much larger range of energies than Cerenkov detectors: from
lower energy photons to photons of hundreds of TeV and more (limited only by
the source), allowing, in principle, to fill the gap between EGRET and Cerenkov
detectors.

\centerline{Acknoledgements}

The authors thank the IFUSP and IFT for providing us with computer resources. 
RAV thanks the IFT for its kind hospitality. This work was supported by CNPq
(COE and PLS) and FAPESP (RAV).

\begin{figure}
\epsfxsize=10cm
\begin{center}
\mbox{\epsfig{file=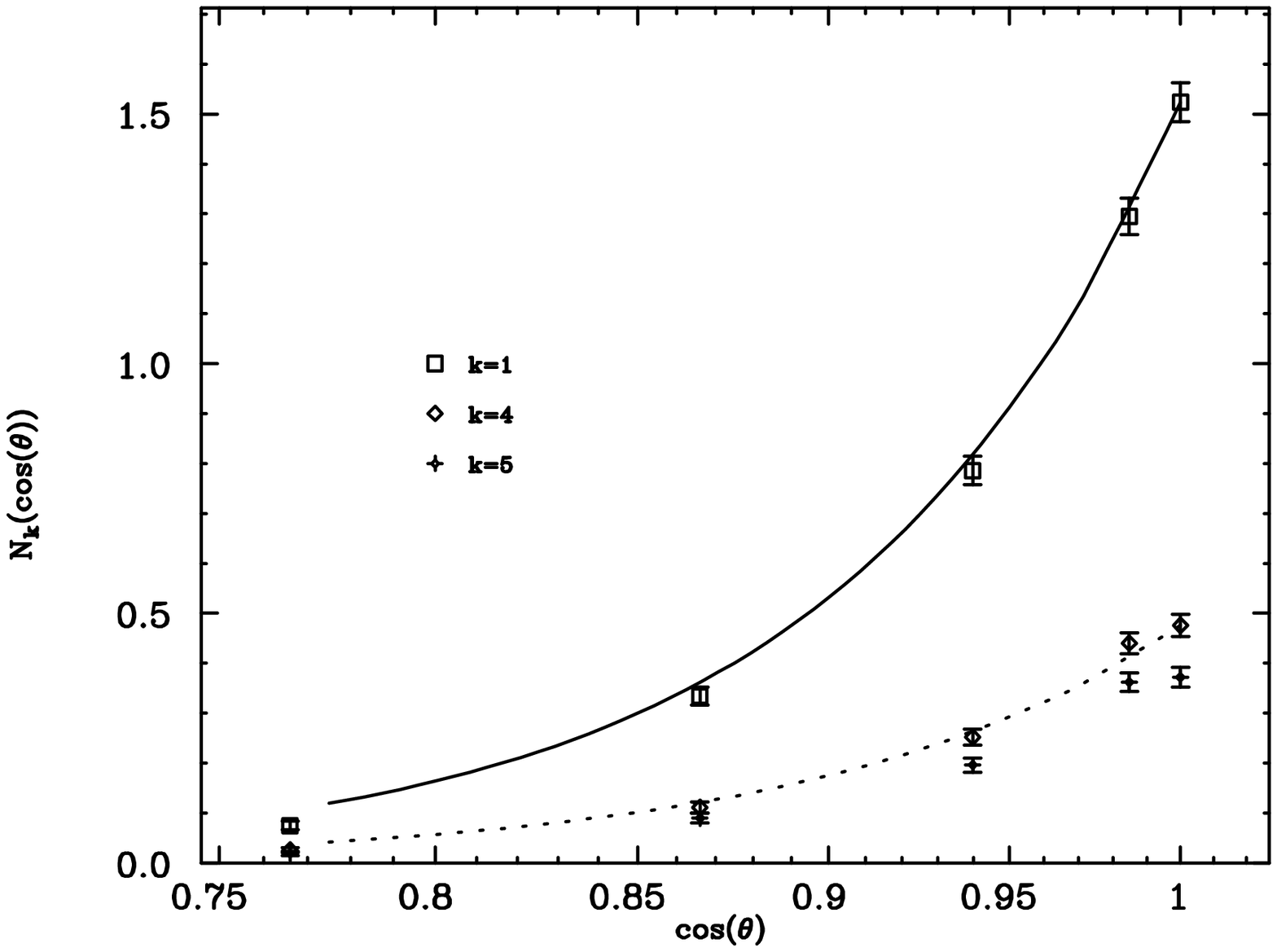}}
\end{center}
\caption{Number of triggers as a function of arrival angle and $k$ for 
photons of 100 GeV. Also shown are the $\cos(\theta)^\alpha$ fit.}
\label{fig2}
\end{figure}
\end{document}